\documentclass[aps,prd,showpacs,floatfix,nofootinbib,superscriptaddress,twocolumn,10pt]{revtex4}
\usepackage{graphicx}
\usepackage{bm}
\usepackage{subfigure}
\usepackage{amssymb}
\usepackage{color,soul}
\usepackage{graphicx}
\usepackage{multirow}
\graphicspath{{figure/}}
\DeclareGraphicsExtensions{.pdf,.png,.jpg}
\setulcolor{red}

\newcommand{\beq}{\begin{equation}}
\newcommand{\eeq}{\end{equation}}
\newcommand{\beqa}{\begin{eqnarray}}
\newcommand{\eeqa}{\end{eqnarray}}

\begin{document}

\title{\ \ \ \ \   XENON1T implications for the DAMA annual modulation:
      \ \ \ \ \ \ \ \ \ \ \ \ \ \ \ \ \ \ \ \ \ \ \ \ \ \ \ \ \ The good, the bad and the ugly.}

\author{R. Foot \footnote {rfoot@unimelb.edu.au}}
\affiliation{School of Physics, University of Melbourne, Victoria 3010 Australia}
\affiliation{School of Physics, University of Sydney, NSW 2006, Australia}

\vskip 5cm

\begin{abstract}

Previous work has argued that the DAMA annual modulation signal might be due to electron recoils in plasma dark matter models.
A specific model
assuming mirror dark matter and featuring collisional shielding of a detector 
due to Earth-bound dark matter was put forward by this author
[arXiv:1806.04293].
That explanation predicted a low energy electron recoil signal in the XENON1T experiment. In light of 
the excess of electron recoils observed in XENON1T, we re-examine
that model's predictions for both DAMA and XENON1T.


\end{abstract}

\maketitle
\vspace{0.6cm}

\section{Introduction}

Dark matter direct detection experiments have the noble aim to discover the interactions of dark matter.
There have been several notable results, including an annual modulation signal
(in the 1-6 keV energy range) observed in the DAMA
detector  \cite{dama1,dama2,dama3}, and more recently,
a low energy excess of electron recoil events in the XENON1T experiment at around $2$ keV\cite{xenon1t}. An
electron recoil excess has also been seen in the Darkside-50 experiment \cite{darkside} at much
lower energy $\sim 0.1$ keV. That excess was interpreted as unmodelled background by the Darkside Collaboration, 
and considered in isolation is not particularly
compelling,
but is relevant to the new physics interpretation of DAMA and XENON1T to be discussed here.
It is also notable that the constraints on dark matter interactions with ordinary matter have reached 
impressive levels, e.g. \cite{Akerib,Cui,aprile}, including  rather stringent limits on electron recoils at low energies \cite{luxmirror,uup,ar,s2only}.

It has been argued in \cite{shield1,shield2} (see also \cite{plasmadm,fj,footyyy} for earlier related work),
that a viable explanation for the DAMA annual modulation is possible (circa mid-2018) in the context of mirror dark matter, and plasma dark matter models
more generally. 
The DAMA signal is interpreted in terms of electron recoils
arising from kinetic mixing induced Coulomb interactions of MeV mass-scale dark electrons. Shielding of the detector from
the halo wind due to Earth-bound dark matter plays an important role.
It can lead to large annual modulation amplitude, as well as an interaction rate with characteristic form:
$dR/dE_R \sim 1/E_R^2$ for $E_R \lesssim E_T$
and more sharply falling for $E_R \gtrsim E_T$. 
The location of the spectral knee, $E_T$, is model dependent.


In this paper we focus on the specific scenario of Ref.\cite{shield1} which assumes mirror dark matter and
features collisional shielding of a detector from the halo wind.
 In the context of that model, we  study here the compatibility of the DAMA
annual modulation with the XENON1T excess and examine other recent constraints.
Although we focus on a specific model, the important features of the interaction rate are
 anticipated to arise quite generally in plasma dark matter models
featuring shielding.
The simplified shielding model of
Ref.\cite{shield1} provides a specific
realization which can be thought of as a useful proxy for a wider range of shielding models, and also particle 
physics models.

This paper is organized as follows. In section II we give a brief discussion of the
most relevant features of the model, and we re-examine the DAMA annual modulation signal for a particular benchmark point suggested in \cite{shield1}.
The predicted shape of the annual modulation spectrum is much steeper than the DAMA measurements - a characteristic feature
of this kind of model with kinetic mixing induced Coulomb interactions. Energy scale and resolution uncertainties
are explored as a means of  `flattening the curve'.
In section III the implications for the XENON1T experiment are examined for the same model and parameter point. In section IV we discuss
the challenge to this interpretation of the DAMA annual modulation signal arising from the XENON1T S2-only analysis.
In section V we conclude.


\section{The DAMA experiment}

For over a decade the DAMA collaboration has observed an annual modulation in a Sodium Iodide detector located at Gran 
Sasso \cite{dama1,dama2,dama3}. 
The modulation has a high statistical significance ($\sim 12 \sigma$) and
is compatible with general expectations from halo dark matter in the Milky Way \cite{sperg}.
A nuclear recoil interpretation is now strongly disfavoured by other experiments, e.g. \cite{ex1,ex2,ex3}. An electron recoil interpretation
is less constrained, although
not without significant obstacles as we will see.

Electron recoils in the keV range are expected to be quite suppressed in the commonly considered WIMP dark matter models, and 
do not appear to be a viable explanation of the DAMA signal \cite{e1,e2,e3}.
However, keV scale electron recoils can
arise quite naturally in plasma dark matter models, the protype considers a dark matter sector consisting of light `dark electrons' and
heavier `dark protons' coupled together via massless `dark photons' \cite{vagnozzi}. 
The dark photon can kinetically mix with the ordinary photon \cite{he}, and thereby
induce a tiny ordinary electric charge of $\epsilon e$ for the dark electron \cite{holdom}. If 
the dark matter halo of the Milky Way takes the form of a highly ionized dark plasma then
energy equipartian implies that dark electrons and dark protons have the same temperature, which
can be in the keV range (or close to keV scale). In that case,  keV electron recoils can result from the Coulomb scattering of (MeV mass scale) 
dark electrons off atomic electrons \cite{plasmadm,footyyy}.

The theoretically most constrained model of this type, mirror dark matter, arises when the dark sector is 
exactly isomorphic to the standard model \cite{flv}.
In that case each ordinary particle has a mirror counterpart of the same mass. 
In that picture, the mirror nuclei and mirror electrons
constitute the inferred dark matter in the Universe. The model is consistent with the large-scale structure constraints \cite{ber,ig,cmb}, early Universe
cosmology \cite{glashow,c19},
and potentially also small-scale structure and galaxy-scale observations (including rotation curves and disk alignment of satellites) \cite{s1,fr,s2h}.
The kinetic mixing interaction plays an important role in that dynamics, leading to 
an estimate of around $\epsilon \sim 10^{-10}$.

For the Milky Way galaxy, the dark halo is modelled as a highly ionized plasma consisting of 
mirror nuclei and mirror electrons, with temperature estimated to be
around $T \approx 0.3$ keV \cite{fv} (for a review see e.g. \cite{freview}).
The nuclear components are
made up of mirror hydrogen and helium nuclei, the latter expected to be the dominant component \cite{ber,paulo}, 
with possibly only trace amounts of heavier elements.

The mirror electrons and mirror nuclei can be detected via scattering on electrons and nuclei. Here, the focus is on electron recoils.
The kinetic mixing induced Coulomb scattering cross section for mirror electrons of velocity, $v$,  scattering on electrons at rest is
\begin{equation}
\frac{d\sigma}{dE_R}=\frac{\lambda}{E_R^2|v|^2}
\label{crosssection}
\end{equation}
where $\lambda = 2\pi\epsilon^2\alpha^2/m_e$.
The cross section on an atom is approximated with a step function, $g_T$, which accounts for the number of bound electrons with binding
energy less than $E_R$.
The rate of electron recoils is then:
\begin{eqnarray}
\label{differential}
\frac{dR}{dE_R} &=& g_T N_{T}n_{e'}\int_{|v|>v_{min}} \frac{d\sigma}{dE_R} f_{e'}(v) |v|d^3v  \nonumber\\
&=& g_T N_{T}n_{e'} \frac{\lambda}{E_R^2} I
\end{eqnarray}
where $N_T$ is the number of target atoms per unit mass and
$f_{e'}(v)$ [$n_{e'}$] is the distribution [number density] of mirror electrons arriving at the detector. The rate is proportional
to the average of the inverse velocity:
\begin{eqnarray}
I \equiv \int_{|v|>v_{min}}^\infty \frac{f_{e'}(v)}{|v|}d^3v 
\label{In}
\end{eqnarray}
with $v_{min} = \sqrt{2E_R/m_e}$.\footnote{Units with $\hbar = c = 1$ are assumed throughout.}
The distribution, $f_{e'}(v)$, even if Maxwellian far from the Earth, is expected to be drastically
modified by shielding effects, as we show now discuss.

Dark matter can accumulate within the Earth, where it thermalizes with the
ordinary matter and forms an extended distribution. This Earth-bound dark matter is expected 
to strongly influence halo dark matter particles near the Earth's surface, effectively shielding the detector
from the halo wind. The mechanism is quite complex involving the generation of dark electromagnetic fields in 
both the Earth-bound dark matter and the halo plasma near the Earth. In addition, collisional shielding can also play
a role.
In fact, it has been argued that these shielding effects can strongly enhance the annual and diurnal modulations, with near maximal annual 
modulation possible \cite{plasmadm,shield1,shield2}.


The shielding is assumed to modify the velocity distribution,
so that only the high velocity tail of the mirror electron distribution survives to reach the detector, assumed here to be
located in the northern hemisphere.\footnote{For a detector located in the southern hemisphere, shielding will stop all halo
dark matter particles from reaching the detector for part of the day, leading to maximal sidereal daily modulation \cite{foots}.}
That is, shielding strongly suppresses $f_{e'}(v)$ below some cut-off, $v_c$.
If this happens, and it is by no means certain, then remarkable simplification occurs:
the velocity integral, $I$ [Eq.(\ref{In})], becomes independent of recoil energy at sufficiently low recoil energies where 
$v_{min} (E_R) = \sqrt{2E_R/m_e} < v_c$.
In that energy range, the energy
dependence of the rate follows that of the cross section. For kinetic mixing induced Coulomb interactions, we
expect the behaviour:
\footnote{Technically, the rate only follows Eq.(\ref{beh}) for $E_R \gtrsim 0.05$ keV, as the cross section at
very low recoil energies 
becomes suppressed due to the lack of loosely bound atomic electrons with binding energies much less than $E_R$. At such low
energies a more sophisticated treatment of the cross section would be required.}
\begin{eqnarray}
dR/dE_R \propto 1/E_R^2\  \  {\rm for} \ E_R \lesssim E_T
\label{beh}
\end{eqnarray}
where $E_T \sim m_e v_c^2/2$.\footnote{In more general plasma dark models, where the dark electron mass
is different from the electron mass, $E_T \sim 2\mu^2 v_c^2/m_e$, where $\mu = m_{e_d} m_e/(m_{e_d}+m_e)$ is the reduced mass.}
For $E_R \gtrsim E_T$ the rate falls much more steeply.

\begin{figure}[htbp]
\centering
\includegraphics[width=1\columnwidth]{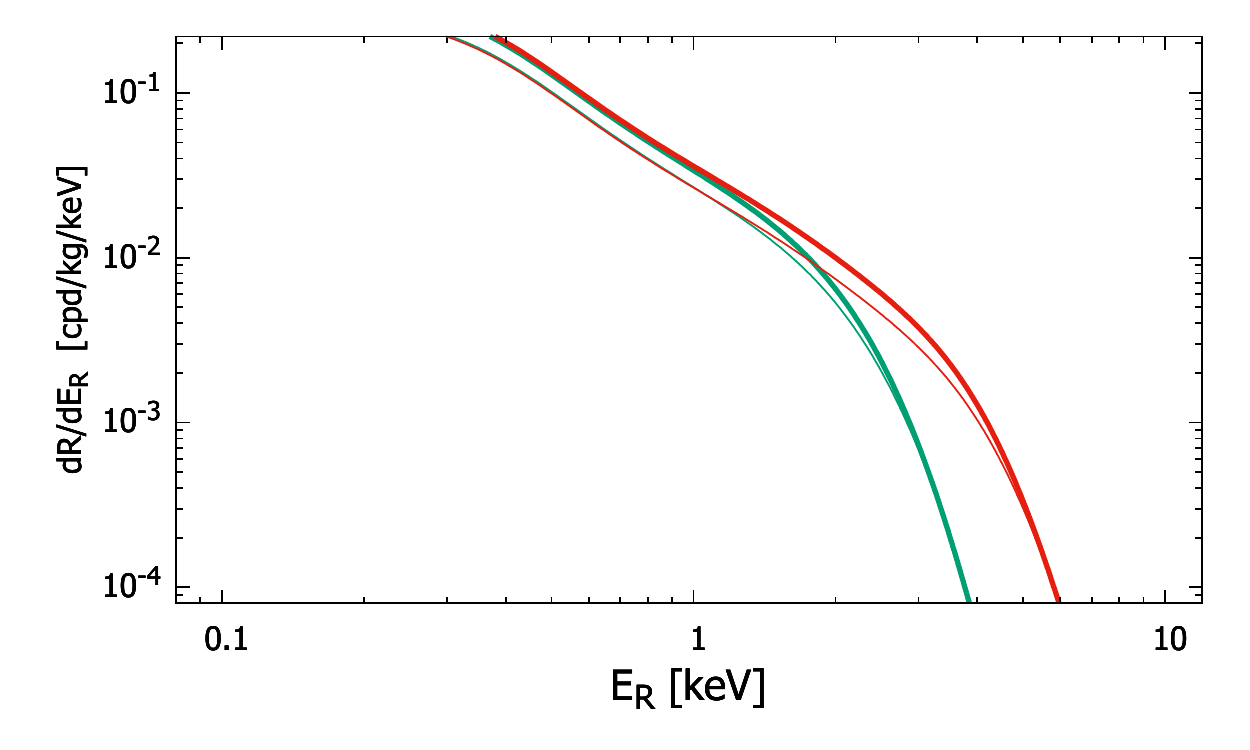}
\vskip -0.45cm
\caption{The model predictions for the
average rate (thick line) and annual modulation (thin line) in the DAMA experiment 
for parameters (a) $T_0 = 0.3$ keV, $T_v = 0.05$ keV (green curves)
and (b) $T_0 = 0.6$ keV, $T_v = 0.1$ keV (red curves).}
\label{fig1}
\end{figure}

\begin{figure}[htbp]
\centering
\includegraphics[width=1\columnwidth]{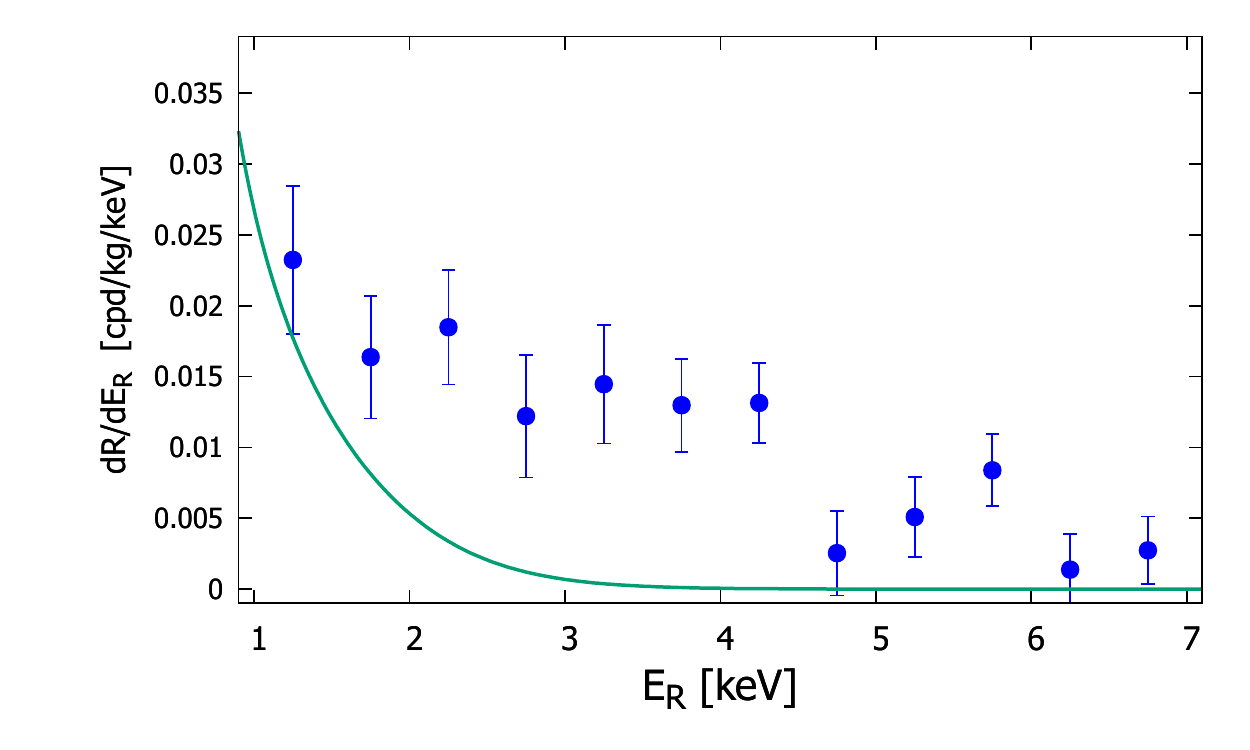}
\vskip -0.45cm
\caption{The model prediction for the annual modulation in the DAMA experiment for the parameter choice (a) compared with 
DAMA/LIBRA phase 2 data from \cite{dama1}.}
\label{fig2}
\end{figure}

In ref.\cite{shield1} we focused on the mirror dark matter model, 
and considered a highly simplified model for collisional shielding of the detector from
Earth-bound mirror helium.
We will not review the details of that shielding model here.
In a sense the details are not so important since, as argued above,
we expect the behaviour to occur quite generally, and may well apply to the case 
where shielding is dominated by dark electromagnetic fields (within the Earth and nearby halo plasma).
In this vain, the simplified shielding model of
Ref.\cite{shield1} provides a specific
realization which can be thought of as a useful proxy for a wider range of shielding models, and also particle physics models.

The model of Ref.\cite{shield1}, hereafter denoted as `the model', was 
defined in terms of four parameters: $\epsilon, n_{\rm He'}(R_E), T_0, T_v$,
where $\epsilon$ is the kinetic mixing strength, $n_{\rm He'}(R_E)$ is the Earth-bound mirror helium density at the Earth's surface,
and $T_0$ ($T_v$)  is the average halo temperature (halo temperature annual variation amplitude) at
the detector location. The analysis of Ref.\cite{shield1} considered two illustrative choices
for $T_0, T_v$, set $\epsilon = 2\times 10^{-10}$, and fixed $n_{\rm He'}$ so that the predicted electron recoil
rate matched the excess observed in the Darkside-50 experiment \cite{darkside} (see below for further explanation).
The two parameter choices considered were (a) $T_0 = 0.3$ keV, $T_v = 0.05$ keV and
(b) $T_0 = 0.6$ keV, $T_v = 0.1$ keV.
The model predictions with these parameter choices for the electron recoil rate in the 
 DAMA/LIBRA experiment
were given in
figure 4a of \cite{shield1}, and for convenience is reproduced here in {\bf Figure 1}.

Figure 1 demonstrates the expected behaviour of the recoil rate [Eq.(\ref{beh})]. From the figure we can
identify the `knee' in the spectrum at
$E_T \approx 1.7$ keV for point (a) and  $E_T \approx 3.5$ keV for point (b).
As mentioned above, the average rate was normalized at low energies to be in rough agreement
with the Darkside-50 excess (at $E_R \sim 0.1$ keV).  A few words are required to explain
the thinking here.
The Darkside Collaboration interpreted their excess as unmodelled background, 
and considered in isolation is not of particular significance.
However, the interaction rate predicted in this model, normalized to the DAMA data at $E_R \sim 1.5$ keV, implies an electron
recoil rate roughly matching the excess observed in the Darkside-50 experiment. That is, the bulk of the Darkside-50 excess 
would have to be due to dark matter interactions in this interpretation of DAMA.
Furthermore, the two experiments can only be consistent with each other if
the annual modulation amplitude is large ($\gtrsim 50\%$), 
which is the case in the specific model considered here. See \cite{shield2} for further discussion.

The predicted rates in Figure 1 were derived by modelling the detector energy resolution in the usual way:
\begin{equation}
\frac{dR}{dE_m}= \int \frac{dR}{dE_R}\frac{e^{-(E_R-E_m)^2/2\sigma^2}}{\sigma \sqrt{2\pi}} \times \epsilon_F (E_R) dE_R
\label{final}
\end{equation}
where $E_m$ denotes the measured recoil energy in the experiment and $\sigma$ is the detector averaged resolution.  
(The efficiency, $\epsilon_F$,
is set to
unity for the DAMA experiment as they give their amplitudes corrected for the efficiency.)
The energy resolution is energy dependent and the standard fit was assumed:
\begin{equation}
\frac{\sigma}{E_R} = \beta + \alpha E_R^{-1/2}
\label{ext8}
\end{equation}
with central values $\alpha = 0.448$ and $\beta = 0.0091$ \cite{damares}.

Henceforth we focus on the parameter choice (a) for further study as the point (b) is disfavoured 
by the scale of the excess observed in XENON1T. 
In {\bf Figure 2}, we compare the predicted modulation amplitude of 
the model for that parameter point with the DAMA/LIBRA phase 2 data \cite{dama1}. 
Evidently, the measured rate is much flatter than the steep spectrum predicted in the model, and 
a very poor description of the data results. Clearly, we need to `flatten the curve'.

\begin{figure}[htbp]
\centering
\includegraphics[width=1\columnwidth]{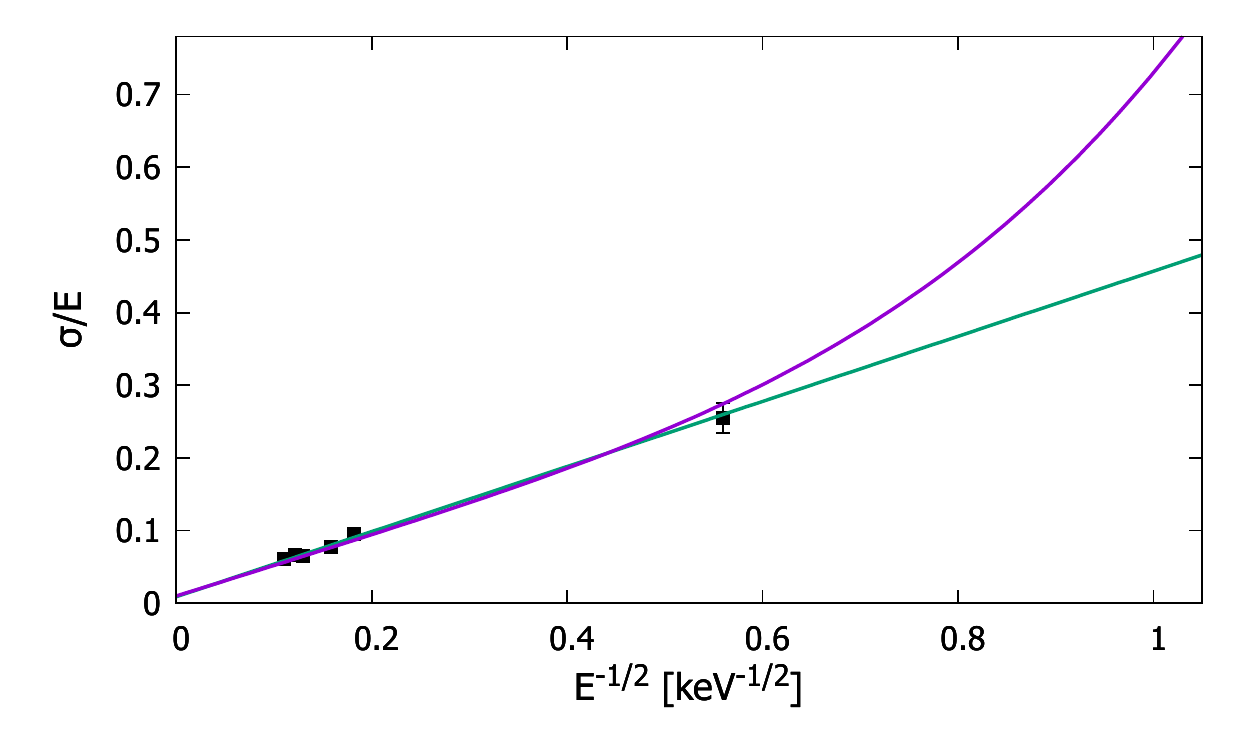}
\vskip -0.45cm
\caption{Comparison of the modified energy resolution case (purple curve) [Eq.(\ref{res})] with the measurements from \cite{damares}.
The green curve is the standard linear model.}
\label{fig3}
\end{figure}

\begin{figure}[htbp]
\centering
\includegraphics[width=1\columnwidth]{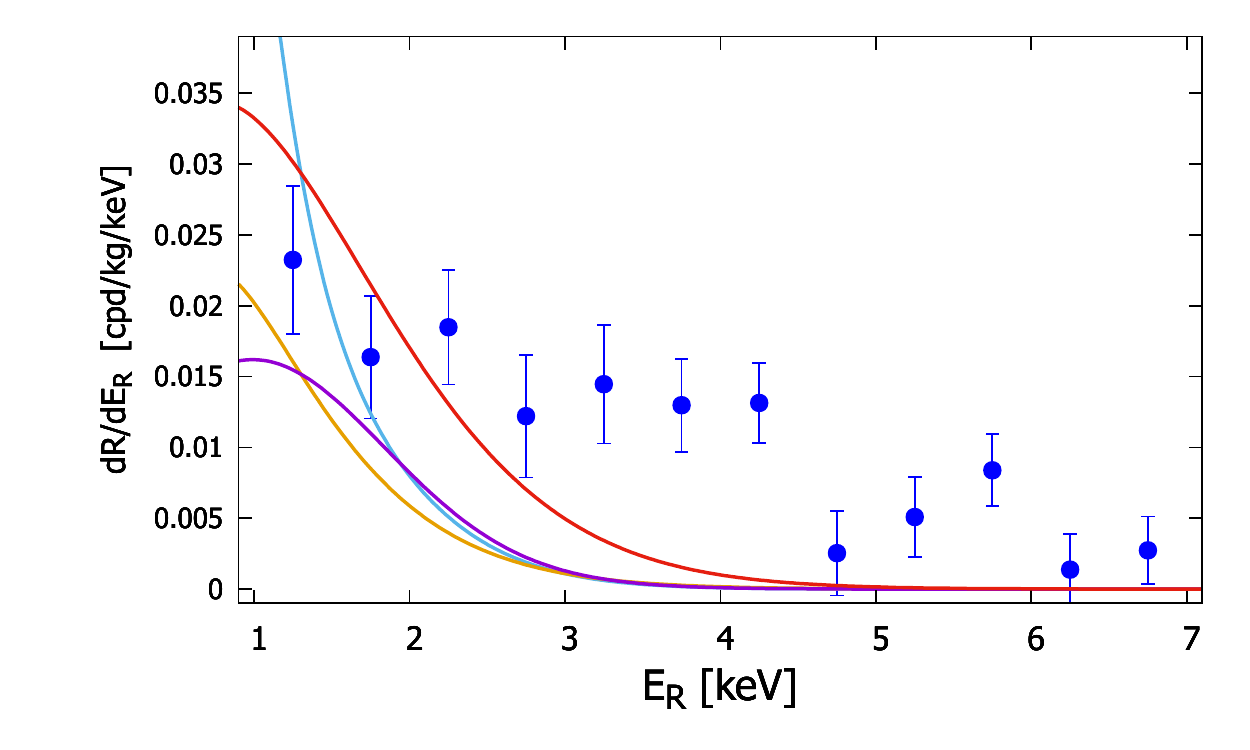}
\vskip -0.45cm
\caption{The model prediction for the DAMA annual modulation with 
(a) small energy scale offset [Eq.(\ref{d1x})] (blue curve),
(b) non-Gaussian resolution behaviour [Eq.(\ref{d2x})] (orange curve), (c) modification of $\sigma/E_R$
with a quartic term [Eq.(\ref{res})] (purple curve), and the red curve shows the effect of all three modifications combined.
}
\label{fig4}
\end{figure}

The situation could possibly be rectified, at least partially, due to uncertainties in the low 
energy response of the detector.
We explore these uncertainties, in a purely phenomenological way, by considering
a small adjustment to the energy scale at low energies and some modifications to the resolution.
The energy scale adjustment considered is
\begin{eqnarray}
E^{\rm true} = E_R - \frac{0.20 \ {\rm keV}}{\sqrt{E_R/{\rm keV}}}
\label{d1x}
\end{eqnarray}
which produces a $\sim 0.1$ keV offset at $E_R = 3.2$ keV, rising to $0.2$ keV offset at $E_R = 1$ keV.
In practice this means that for each $E_R$ value, the rate in the integrand in Eq.(\ref{final}) is evaluated at the true energy,
as given by the above equation, with the Jacobian factor, $J = 1 + 0.10(E_R/{\rm keV})^{-3/2}$ incorporated. For the standard resolution
case, no changes are made to the way resolution is incorporated, i.e. $\sigma$ is defined in terms of $E_R$, not $E^{\rm true}$, and 
is given in Eq.(\ref{ext8}). 

With regards to the resolution modification, we first examine the effect of a deviation away from Gaussian behaviour.
To examine such an effect, we make the replacement in Eq.(\ref{final}) of
\begin{eqnarray}
\frac{e^{-(E_R - E_m)^2/2\sigma^2}}{\sigma \sqrt{2\pi}} \to \frac{e^{-(E_R-E_m)^{p}/2\sigma^{p}}}{\sigma {\cal N}(p) \sqrt{2\pi}}\ .
\label{d2x}
\end{eqnarray}
For $p < 2$ ($p > 2$) the distribution is enhanced (suppressed) in the tail compared to the standard Gaussian distribution.
Since we are interested in flattening the curve, we shall consider the $p < 2$ case, choosing for definiteness
$p = 1.5$. The corresponding normalization factor is ${\cal N} \simeq 1.143$.
Finally, the size of the width, $\sigma$, could potentially deviate at low energies from the linear extrapolation, Eq.(\ref{ext8}), where
measurements are unavailable.
Specifically, we examine the inclusion of a quartic  term (in $E_R^{-1/2}$):
\begin{eqnarray}
\frac{\sigma}{E_R} = \beta + \alpha E_R^{-1/2}  + \gamma E_R^{-2} 
\label{res}
\end{eqnarray}
with  $\beta = 0.01$, $\alpha = 0.42$, $\gamma = 0.30$. The comparison of this
resolution function with the calibration data is shown in {\bf Figure 3}.

The rising $\sigma$ values might unduly enhance the predicted rate above threshold due to upward 
fluctuations of very low energy events ($E_R \lesssim 0.5$ keV).
Such an effect could not be realistic due to sharply falling efficiency in this very low energy region.
In analyzing these non-standard cases, we therefore assume
a low energy cut due to loss of efficiency, $\epsilon_F (E_R) = 0$ for $E_R \le 0.5$ keV, and $\epsilon_F (E_R) = 1$ for $E_R > 0.5$ keV. 
Irrespective of the resolution modification, such a cut should make the modelling of the low energy response in the detector more realistic.
Naturally, it would be preferable to deal with data uncorrected for efficiency and incorporate the efficiency function estimated from calibration data. 
However, the
information provided by the DAMA collaboration is insufficient for this to be done.

{\bf Figure 4}  shows the effect of the energy scale and resolution modifications.
That figure shows some improvement, and one might take the optimistic view that there is some potential
to explain the DAMA/LIBRA measurements below 3 keV.
However, the existence of substantial annual modulation above 3 keV seems quite puzzling from this perspective.
It suggests more drastic modifications to the ones considered here, whether such significant modifications
are possible is, of course, unclear.


The flattening the curve issue is not the only thing to worry about.
This interpretation of the DAMA annual modulation appears to be at odds with the results from the LUX experiment.
That experiment did not find any evidence for electron recoils in excess of their background model,
and their study \cite{luxmirror} excluded the model discussed here. More generally, they have obtained very
impressive constraints on annual and diurnal modulations of electron recoils \cite{uup}. However, there are a couple of
loose ends, one of which is the location issue.
The LUX detector operated in the Homestake mine, while DAMA, XENON1T and Darkside-50 are all located
in Gran Sasso. While both locations have similar latitude ($44.3^{\circ}$ for Homestake, $42.4^{\circ}$ for Gran Sasso),
the surrounding geography is very different. In fact,
the Gran Sasso location seems somewhat special as it is high in altitude relative to the surrounding terrain.
This relatively high altitude might reduce the influence of shielding, and even a modest reduction could be important as the 
interaction rate is exponentially 
sensitive to such variations, given that only the high velocity tail of the mirror electron distribution survives to reach the detector.
Although rough calculations find that such effects are unlikely to give more than a `factor of two' difference in electron recoil rate in the collisional
shielding model considered, the realistic case is anticipated to be more complex and there is potential for more important location
dependence.
\footnote{Generally, dark electromagnetic fields may play an important role in the shielding mechanism. In particular, halo dark
matter can ionize a surface layer of the Earth-bound mirror helium leading to the formation of a `dark ionosphere' \cite{plasmadm}.
In that case, the importance of shielding is expected to be quite sensitive to the height of the detector relative to the
dark ionosphere. If the dark ionosphere is located close to the Earth's surface, then significant location dependence seems possible,
even for detectors at the same latitude.}

\section{The XENON1T experiment}

The XENON1T experiment has observed an excess of electron recoil events in a combined S1 and S2 analysis near their low
energy threshold of around 2 keV \cite{xenon1t}. The statistical significance of the excess
is modest, $\sim 3.5 \sigma$, and while background interpretations exist \cite{xenon1t,pole}
one can also interpret the excess as a possible signal for new physics \cite{xenon1t}. Indeed,
the XENON1T excess has sparked much interest
in the theoretical community, including new physics models involving axions, nonstandard neutrino properties as well as dark matter 
e.g. \cite{range} (by no means an exhaustive list of the many proposals offered).

The XENON1T excess is compatible with a rising $dR/dE_R \propto 1/E_R^2$ rate at low energies, except possibly the lowest
energy bin, and interpretations in terms of mirror dark matter \cite{md1}, and more general dark plasma models \cite{md2}
have already been proposed.
The first of these papers examined mirror dark matter with shielding treated in a general way, as in Eq.(\ref{beh}), and
considered parameter space where $E_T \ge 4$ keV. The second paper, which included this author, looked at the more
general plasma dark matter case, and studied the parameter region where the Earth-bound dark matter is so compact
as to make shielding unimportant.
These papers, as they focused solely on XENON1T, do not address the pertinent question of 
the compatibility of the XENON1T excess with the DAMA annual modulation signal.


\begin{figure}[t]
\centering
\includegraphics[width=1\columnwidth]{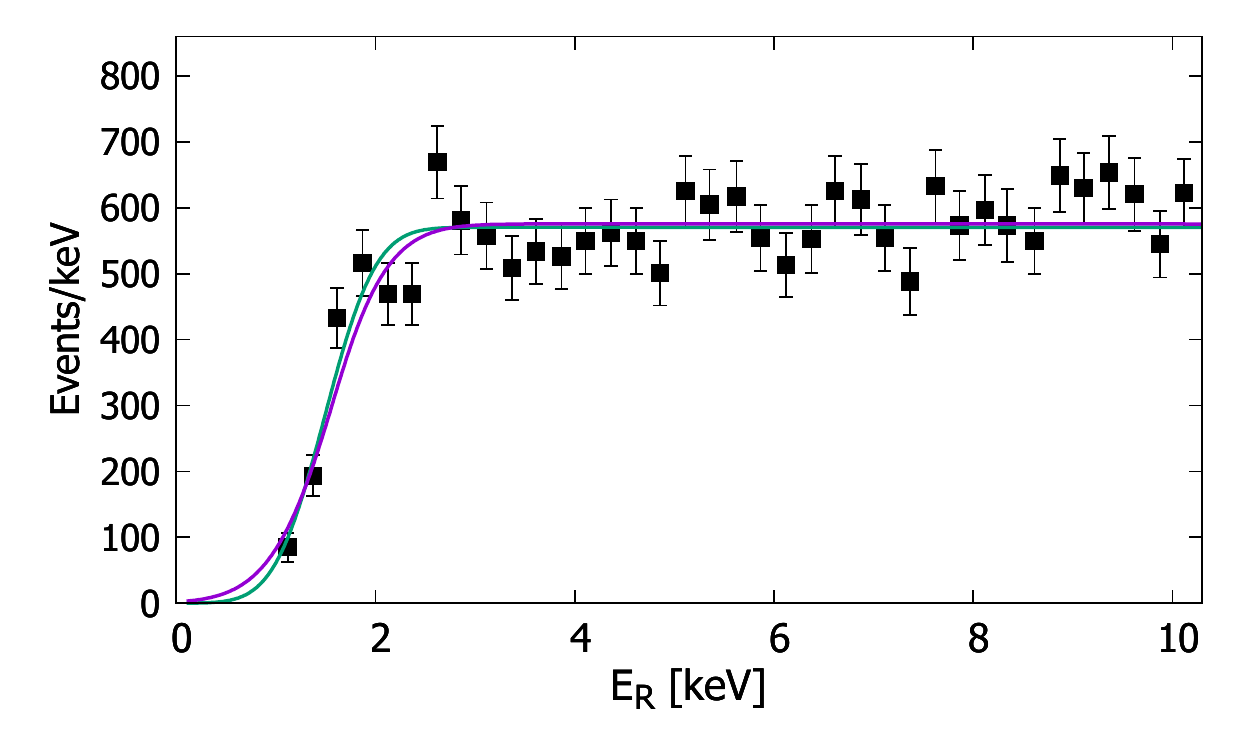}
\vskip -0.45cm
\caption{Modelling the $^{220}$Rn calibration data. The green curve shows the best fit with standard
resolution assumed,
while the purple curve shows the best fit assuming modified resolution (see text).}

\label{fig5}
\end{figure}

\begin{figure}[t]
\centering
\includegraphics[width=1\columnwidth]{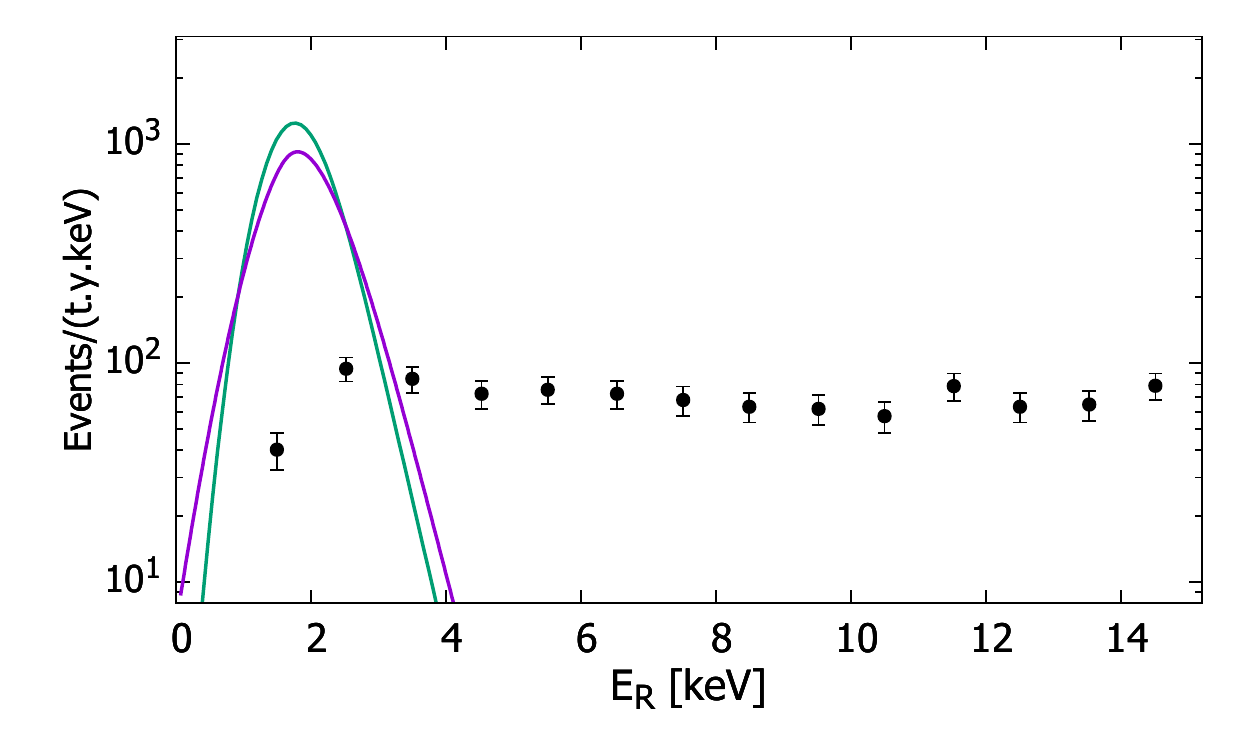}
\vskip -0.4cm
\caption{The model prediction for the low energy recoil rate in XENON1T. 
The  green curve assumes standard resolution.
while the purple curve assumes modified resolution.}
\label{fig6}
\end{figure}

Returning to the particular model discussed in section II,
it was noted in Ref.\cite{shield1} that the model predicts an electron recoil average rate for the XENON1T experiment
 in the  2-6 keV range of $\sim 10^{-4}$ cpd/kg/keV for the parameter point (a) discussed here.
That estimate assumed a sharp cutoff in the detection
efficiency of the S1 signal conservatively taken to be 2.2 keV. The calibration data given in \cite{xenon1t} suggests that this
is somewhat higher ($\sim 0.6$ keV) than realistic, however there can be energy scale uncertainties. Our purpose here
is to study the predicted rate in more detail and compare with the XENON1T data.

The rate of electron recoils predicted in the XENON1T experiment takes the 
general form as given in Eq.(\ref{final}). There is a modest adjustment to the cross section arising from a slight change to the
number of loosely bound atomic electrons at 2 keV (44 for Xe, 54 for NaI).
The modelling of the efficiency and resolution are of central importance.

The energy resolution for the XENON1T experiment can be modelled with a Gaussian distribution with $\sigma$ of the form, Eq.(\ref{ext8}), 
with $\alpha = 0.31$, $\beta = 0.0037$ \cite{xenon1t,xenonres}. In the analysis presented here, we also consider a nonstandard resolution case,
as in Eq.(\ref{d2x}), setting $p = 1.6$ (${\cal N} \simeq 1.103$). Again, this is a purely phenomenological modification employed as a means to
flatten the curve, not otherwise justified; there are better motivated modifications, 
such as a skewed normal distribution \cite{pole} which could be considered
instead, especially if one is of the view that the ends does  not justify the means.

\begin{figure}[htbp]

\subfigure[\ ]{\includegraphics[width=0.92\columnwidth]{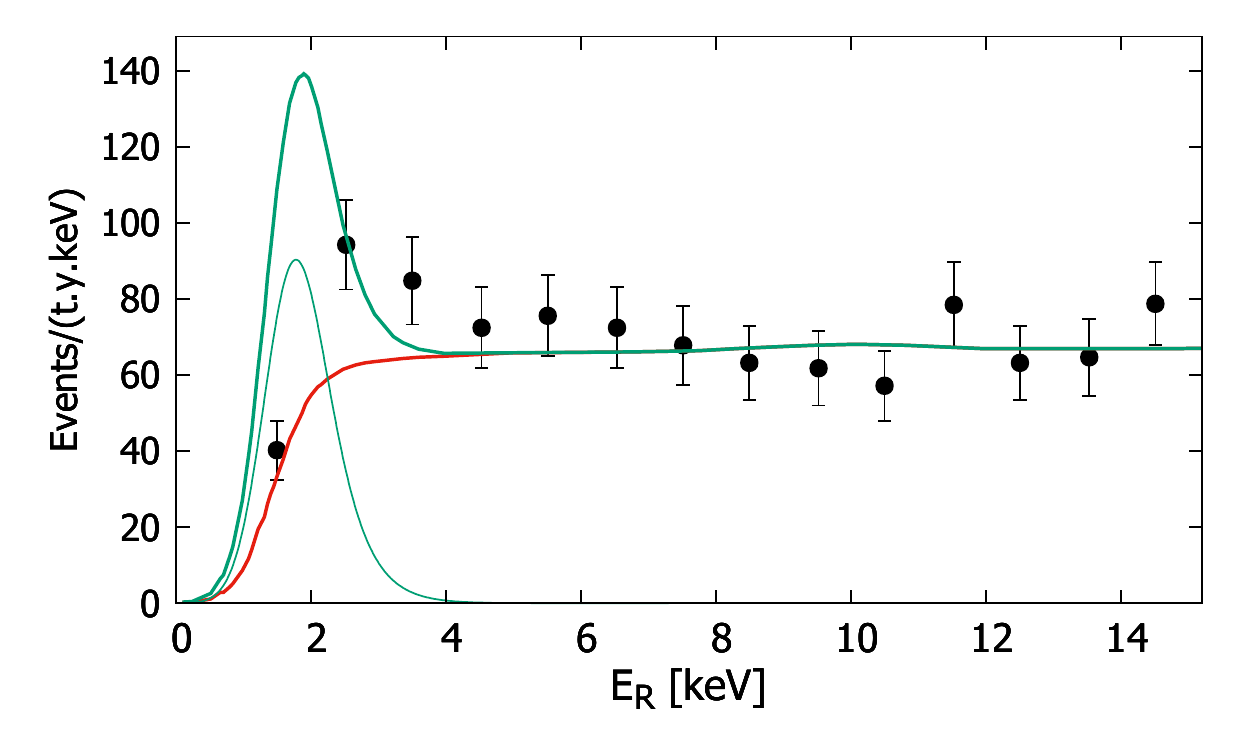}}

\subfigure[\ ]{\includegraphics[width=0.92\columnwidth]{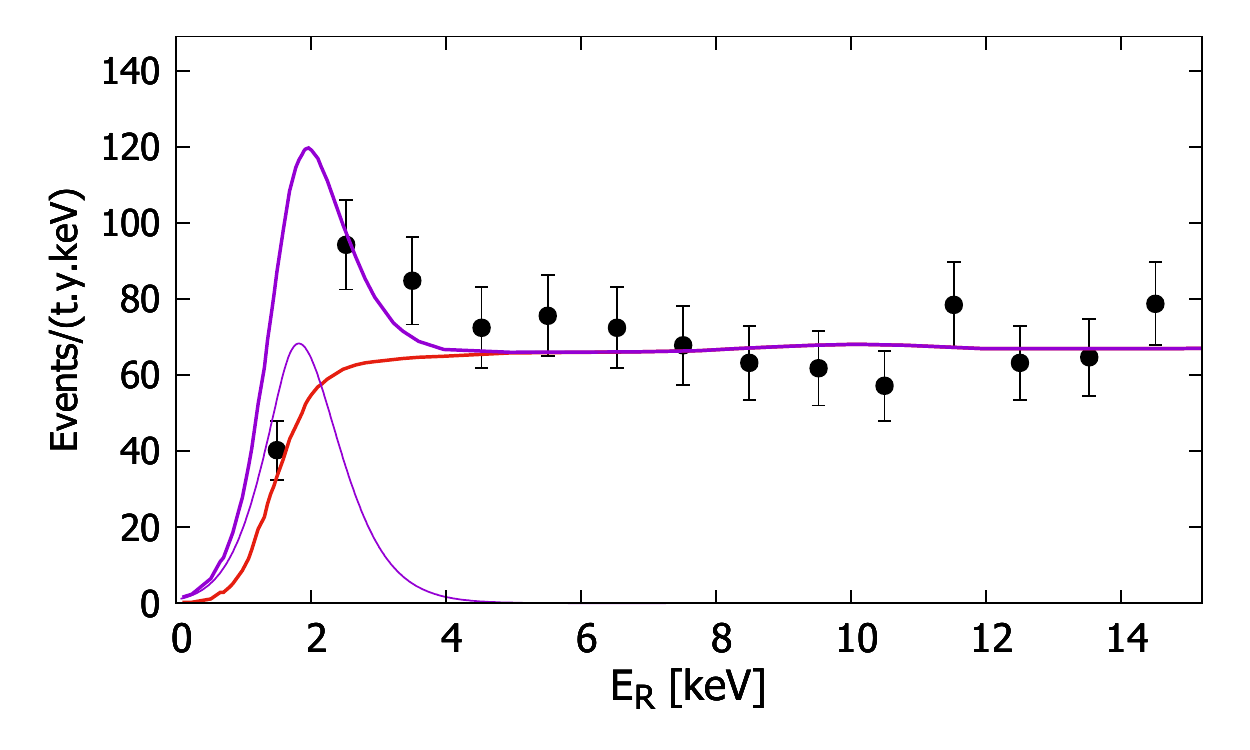}}

\vskip -0.3cm
\caption{
The model prediction for the low energy recoil rate in XENON1T. 
A $\sim 0.6$ keV energy scale correction is 
assumed, Eq.(\ref{mis}). 
(a) assumes standard resolution while (b) is the modified resolution case considered. 
The red curve is the  XENON1T $B_0$ background model.
}

\label{fig7}
\end{figure}

The detector efficiency function, $\epsilon_F(E_R)$, is also very important,
and can be estimated by using
 the $^{220}{\rm Rn}$ calibration data (figure 6 of Ref.\cite{xenon1t}).
Approximating the $^{220}{\rm Rn}$ beta decay spectrum as flat at low energies ($E_R < 10$ keV),
and modelling the efficiency with a sigmoid function (defined with two free parameters $\lambda$, $E_R^0$):
\begin{eqnarray}
S(E_R) = \frac{1}{1 + e^{-\lambda (E_R - E_R^0)}}
\label{sigm}
\end{eqnarray}
we obtain a good fit to the calibration data, shown in {\bf Figure 5}.
A $\chi^2$ analysis, assuming standard resolution,
gives $\chi^2_{\rm min} = 37.1$ for 34 degrees of freedom for the best fit values: $\lambda = 26.2,\ E_R^0 = 1.55$ keV.
This exercise, repeated for the modified resolution case considered, yields
$\chi^2_{\rm min} = 39.8$ for 34 degrees of freedom with best values: $\lambda = 28.1$, $E_R^0 = 1.61$ keV.
Note the efficiency function used includes a normalization factor of 0.87, mainly due to the selection efficiency
cut \cite{xenon1t}. That is, $\epsilon_F (E_R) = 0.87 \times S(E_R)$.

It is straightforward to evaluate the rate for the model [for the parameter point (a) with
$T_0 = 0.3$ keV, $T_v = 0.05$ keV, $\epsilon = 2\times 10^{-10}$].
{\bf Figure 6} shows the result of this computation for both standard resolution and the alternative case.
In both cases this
rate is over an order of magnitude larger than the observed excess. 
This is disappointing, however the tension might not be quite as severe 
as Figure 6 suggests due to the sensitivity of the predictions to energy scale uncertainties.

Since the recoil rate is rising sharply at low energies  [$dR/dE_R \propto 1/E_R^2$,
for $E_R \lesssim E_T$, with a spectral `knee' at $E_T \sim 1.7$ keV], the predicted rate will 
be quite sensitive to a mismodelling of the energy scale. In ref. \cite{pole} it has been argued that
a slight $\sim 0.5$ keV offset is plausible at around $2.3$ keV rising even higher at low energies. That is, the measured 
$2.3$ keV energy could correspond to $2.8$ keV true recoil energy.
To explore the possible effect of such a mismodelling of the energy scale we consider:
\begin{eqnarray}
E_R^{\rm true} = E_R + \frac{0.95\ {\rm keV}}{\sqrt{E_R/{\rm keV}}}
\label{mis}
\end{eqnarray}
which produces a $\sim 0.6$ keV offset at $E_R^m = 2.3$ keV (marginally higher than the $0.5$ keV offset
considered in \cite{pole}) and rising higher at lower energies.
{\bf Figure 7a} (standard resolution)
and {\bf Figure 7b} (modified resolution) show the rate,
 taking into account the effect of a mismodelled energy scale as given in Eq.(\ref{mis}).

The normalization of the predicted excess is still somewhat high (factor of $\sim 2$), which suggests an even larger energy
scale mismodelling or some other inaccuracy in the way the rate is modelled. A reduction in kinetic
mixing strength, $\epsilon$, could also be considered, but such reduction would have to be modest if the 
DAMA signal is to be accommodated.
Still, Figure 7 does suggest that the XENON1T excess might be consistent with the model.
That is, a simultaneous explanation for 
both the XENON1T electron recoil excess and the DAMA annual modulation seems possible, however there are 
qualifications.
With regard to DAMA (as discussed in section II) there are tensions in explaining the overall shape of the
annual modulation, and there are some compatibility issues with LUX constraints.
However, the most serious constraint on this interpretation of the DAMA annual modulation signal arises from 
the XENON1T S2-only analysis.

\section{The elephant in the room: The XENON1T S2 signal}


We have yet to discuss the elephant in the room. The model's increasing rate at low 
energies, $dR/dE_R \propto 1/E_R^2$ suggests a rate substantially larger than the limit claimed by the XENON1T S2-only analysis \cite{s2only}.
That analysis considered the S2-only events and employed an array of cuts with the aim to effectively remove the substantial background.
The cuts are quite severe as the effective exposure of $\sim$ 20 tonne-day (for $E_R \gtrsim 0.3$ keV) is around 20 times
lower than that of XENON1T's main analysis \cite{aprile}.
The XENON1T analysis results in the impressive limit of $dR/dE_R \lesssim 1$ events/tonne/day/keV at $E_R \sim 1$ keV. In comparison
the model discussed here predicts around 30 events/tonne/day/keV at that energy. 

Uncertainties in the energy scale are unlikely to resolve the disagreement.
The total number of electron recoils between 0.3 keV and 3 keV, for a 20 tonne-day exposure is estimated to be around 1200 in the model.
This can be compared with the $\sim $ 50 or so events observed between S2[PE] 150 - 3000.
The conclusion is clear:
{\it the interpretation of the DAMA annual modulation signal discussed in this paper is not compatible with the XENON1T S2-only analysis.}

The cuts employed in the XENON1T S2-only analysis are quite important, and nontrivial,
the most significant of which involves a constraint on the S2 width.
Without a S1 signal, the interaction depth $z$ cannot be determined.
Since the width of the S2 waveform in time is expected to correlate with the depth due to diffusion of the electrons during drift \cite{sor},
the S2 width can be used to help locate events in the interior of the detector where backgrounds are low.
The data when plotted in terms of S2 width and S2 [PE] show a region with very few events, from which very stringent constraints follow
(of order 1000 times more stringent than the S2-only analysis of XENON100 data, where such cuts were not employed \cite{s2100}).
The impressive limits from the XENON1T S2-only analysis assume, therefore, that the correlation of S2 width with depth is calibrated correctly.
The calibration of the S2 width with $z$, was modelled using XENON1T's waveform simulator \cite{abc}, and is in
nice agreement with neutron generator calibration data \cite{supx}, 
although the neutron generated events are typically of higher energy as both S1 and S2 events are observed.

Although the interpretation of the DAMA annual modulation discussed in this paper is incompatible with the XENON1T S2-only
constraints, the incompatibility does not extend to interpretations of the XENON1T excess
on its own, within the class of models discussed here.
For example, without modification to the energy scale, the model's prediction for the XENON1T excess, given in Figure 6, would need
to be scaled by down by a factor of around $\sim$ 30 (since the rate is proportional to $\epsilon^2$, this factor of 
30 reduction can be achieved by setting $\epsilon \simeq 3.7 \times 10^{-11}$ instead of $\epsilon = 2\times 10^{-10}$).
This scaling will reduce the number of events in the 0.3 keV - 2 keV range from around 1200 to about 40 or so for a 20 tonne-year exposure.
This revised rate would presumably be compatible with the S2-only
analysis, when modest energy scale uncertainties are included.  
In particular the plasma dark matter interpretations of the XENON1T excess discussed in \cite{md1} and \cite{md2} are likely compatible
with the XENON1T S2-only analysis.\footnote{In these models, the integrated number of events in the 0.3 keV - 2 keV range for a 20 tonne-year exposure in xenon
would be lower, around 10-20 events, as they model the XENON1T excess without a spectral knee \cite{md2} or one with $E_T > 4$ keV \cite{md1}.}

\section{Discussion}

In this paper we have examined a particular dark matter model with several distinctive
features. These include: Milky Way halo composed of highly ionized dark plasma, shielding of a detector due to Earth-bound dark matter,
and a Coulomb-type interaction with ordinary matter induced by the kinetic mixing interaction.
These features all occur in the mirror dark matter model, where one contemplates a hidden sector which is an exact duplicate of the standard model.
They can also arise in closely related, but more general dark sector models.
Electron recoils in the keV energy range naturally occur in such models due to dark electron Coulomb scattering
off atomic electrons. 

The shielding plays an important role in modifying the distribution of halo dark electrons reaching the detector. 
If only the high velocity tail
of the distribution survives to reach the detector then a simple picture emerges: The
interaction rate falls like $dR/dE_R \propto 1/E_R^2$ for $E_R \lesssim E_T$
and more sharply falling for $E_R \gtrsim E_T$. 
The location of the spectral knee, $E_T$, is model dependent.
A particular shielding model was developed in \cite{shield1}, and two 
parameter points studied, one with $E_T \sim 1.7$ keV and the other with $E_T \sim 3.5$ keV. At that time
it was argued that the DAMA annual modulation signal could be explained consistently with other experiments
for both parameter points, and that XENON1T should provide an important test of the model.
Our initial purpose here was to re-examine that explanation of the DAMA annual modulation signal in light
of the XENON1T results reported in \cite{xenon1t}. We have found that
a simultaneous explanation of the DAMA signal and low energy XENON1T electron recoil excess appears
possible, although some significant tension remains. The spectral shape of the DAMA annual modulation measurements is rather
difficult to accommodate, while the normalization of the XENON1T excess suggests a $\sim 0.6$ keV energy mismodelling
in that experiment. 

The LUX collaboration did not see any electron recoil excess,
and their study in \cite{luxmirror} excluded the parameter point discussed here. More generally, they have obtained very
strong constraints on annual and  daily modulation of electron recoil events \cite{uup}.
However, there are caveats. The LUX detector operated in the Homestake mine, while DAMA, XENON1T and Darkside-50 are all located
in Gran Sasso. The shielding effect could be
sensitive to a variation in height above sea level, elevation of surrounding terrain etc., that is, feature location dependence.
Although such dependence is expected to be modest in the collisional shielding model considered, 
dependence on location could be more significant in the more general case where dark electromagnetic fields play a role in shielding (e.g. if there
is a dark ionosphere near the Earth's surface).
Also, the rate in the LUX experiment should show
the same sensitivity to a small energy scale mismodelling as was discussed here in the
context of XENON1T, where a $\sim 0.6$ keV energy scale recalibration
reduced the rate by over an order of magnitude (figure 6 cf. figure 7a).

An even more serious challenge to the plasma dark matter interpretation of the DAMA modulation arises from the S2-only analysis
of XENON1T data \cite{s2only}.
That analysis obtained a very stringent limit on electron recoils of $dR/dE_R \lesssim 1$ events/tonne/day/keV at $E_R \sim 1$ keV.
 There is clear incompatibility with that analysis and the model discussed here.
If this interpretation of the DAMA annual modulation  is to survive it would require the XENON1T S2-only analysis to 
have some fundamental flaw. 
The incompatibility does {\it not} extend to plasma dark matter interpretations of the XENON1T excess on its own, 
as considered recently in \cite{md1,md2}, as the electron recoil rate at $E_R \sim 1$ keV is estimated 
to be around 0.1-1 events/tonne/day/keV in those scenarios.

The XENON1T S2-only analysis does suggest that  the DAMA annual modulation
explanation given by this author in \cite{shield2} is unlikely. 
Nevertheless, direct detection experiments appear as an ideal way to probe plasma 
dark matter models (and, of course, the general concept of dark matter), and the near
future looks promising with sensitive forthcoming experiments, including XENONnT \cite{pan1}, LZ \cite{pan2}, PandaX \cite{pan3}.
There is some hope, therefore, that
the scientific `Holy Grail', that the discovery of dark matter would
represent, will not remain elusive for much longer.

\vspace{0.1cm}

\noindent
{\bf Acknowledgments}: \ The author would like to thank M.~Szydagis for invaluable correspondence, and for comments on
a draft of this article.

\vskip 0.4cm



\begin{thebibliography}{}

\bibitem{dama1}
R.~Bernabei {\it et al}., 
Nucl. Phys. Atom. Energy \textbf{19}, 307-325 (2018)
[arXiv:1805.10486].

\bibitem{dama2}
R.~Bernabei {\it et al}.,
Eur. Phys. J. C \textbf{73}, 2648 (2013)
[arXiv:1308.5109].


\bibitem{dama3}
R.~Bernabei \textit{et al.} [DAMA and LIBRA],
Eur. Phys. J. C \textbf{67}, 39-49 (2010)
[arXiv:1002.1028].


\bibitem{xenon1t} 
E. Aprile {\it et al.} [XENON], arXiv:2006.09721.


\bibitem{darkside}
P.~Agnes \textit{et al.} [DarkSide],
Phys. Rev. Lett. \textbf{121}, 111303 (2018)
[arXiv:1802.06998].





\bibitem{Akerib}
D.~S.~Akerib \textit{et al.} [LUX],
Phys. Rev. Lett. \textbf{118}, 021303 (2017)
[arXiv:1608.07648].

\bibitem{Cui}
X.~Cui \textit{et al.} [PandaX-II],
Phys. Rev. Lett. \textbf{119}, 181302 (2017)
[arXiv:1708.06917].

\bibitem{aprile}
E.~Aprile \textit{et al.} [XENON],
Phys. Rev. Lett. \textbf{121}, 111302 (2018)
[arXiv:1805.12562].


\bibitem{luxmirror}
D.~S.~Akerib \textit{et al.} [LUX],
Phys. Rev. D \textbf{101}, 012003 (2020)
[arXiv:1908.03479].


\bibitem{uup}
D.~S.~Akerib \textit{et al.} [LUX],
Phys. Rev. D \textbf{98}, 062005 (2018)
[arXiv:1807.07113].


\bibitem{ar}
E.~Aprile \textit{et al.} [XENON100],
Science \textbf{349}, 851-854 (2015)
[arXiv:1507.07747].


\bibitem{s2only}
E.~Aprile \textit{et al.} [XENON],
Phys. Rev. Lett. \textbf{123}, 251801 (2019)
[arXiv:1907.11485].




\bibitem{shield1}
R.~Foot,
Phys. Lett. B \textbf{789}, 592-597 (2019)
[arXiv:1806.04293].


\bibitem{shield2}
R.~Foot,
Phys. Lett. B \textbf{785}, 403-408 (2018)
[arXiv:1804.11018].


\bibitem{plasmadm}
J.~D.~Clarke and R.~Foot,
JCAP \textbf{01}, 029 (2016)
[arXiv:1512.06471].


\bibitem{footyyy}
R.~Foot,
Phys. Rev. D \textbf{90}, 121302 (2014)
[arXiv:1407.4213].

\bibitem{fj}
J.~D.~Clarke and R.~Foot,
Phys. Lett. B \textbf{766}, 29-34 (2017)
[arXiv:1606.09063].

\bibitem{sperg}
A.~K.~Drukier, K.~Freese and D.~N.~Spergel,
Phys. Rev. D \textbf{33}, 3495-3508 (1986).

\bibitem{ex1}
C.~Savage, G.~Gelmini, P.~Gondolo and K.~Freese,
JCAP \textbf{04}, 010 (2009)
[arXiv:0808.3607].

\bibitem{ex2}
C.~Savage, K.~Freese, P.~Gondolo and D.~Spolyar,
JCAP \textbf{09}, 036 (2009)
[arXiv:0901.2713].

\bibitem{ex3}
S.~Kang, S.~Scopel, G.~Tomar, J.~H.~Yoon and P.~Gondolo,
JCAP \textbf{11}, 040 (2018)
[arXiv:1808.04112].

\bibitem{e1}
R.~Bernabei {\it et al}.,
Phys. Rev. D \textbf{77}, 023506 (2008)
[arXiv:0712.0562].

\bibitem{e2}
J.~Kopp, V.~Niro, T.~Schwetz and J.~Zupan,
Phys. Rev. D \textbf{80}, 083502 (2009)
[arXiv:0907.3159].


\bibitem{e3}
B.~M.~Roberts, V.~A.~Dzuba, V.~V.~Flambaum, M.~Pospelov and Y.~V.~Stadnik,
Phys. Rev. D \textbf{93}, 115037 (2016)
[arXiv:1604.04559].


\bibitem{vagnozzi}
R.~Foot and S.~Vagnozzi,
Phys. Rev. D \textbf{91}, 023512 (2015)
[arXiv:1409.7174].

\bibitem{he}
R.~Foot and X.~G.~He,
Phys. Lett. B \textbf{267}, 509-512 (1991).

\bibitem{holdom}
B.~Holdom,
Phys. Lett. B \textbf{166}, 196-198 (1986).



\bibitem{flv}
R.~Foot, H.~Lew and R.~R.~Volkas,
Phys. Lett. B \textbf{272}, 67-70 (1991).

\bibitem{ber}
Z.~Berezhiani, D.~Comelli and F.~L.~Villante,
Phys. Lett. B \textbf{503}, 362-375 (2001)
[arXiv:hep-ph/0008105].


\bibitem{ig}
A.~Y.~Ignatiev and R.~R.~Volkas,
Phys. Rev. D \textbf{68}, 023518 (2003)
[arXiv:hep-ph/0304260].


\bibitem{cmb}
R.~Foot,
Phys. Lett. B \textbf{718}, 745-751 (2013)
[arXiv:1208.6022].



\bibitem{glashow}
E.~D.~Carlson and S.~L.~Glashow,
Phys. Lett. B \textbf{193}, 168-170 (1987).

\bibitem{c19}
R.~Foot,
Phys. Lett. B \textbf{711}, 238-243 (2012)
[arXiv:1111.6366];
P.~Ciarcelluti and R.~Foot,
Phys. Lett. B \textbf{679}, 278-281 (2009)
[arXiv:0809.4438].





\bibitem{s1}
R.~Foot and S.~Vagnozzi,
JCAP \textbf{07}, 013 (2016)
[arXiv:1602.02467].

\bibitem{fr}
R.~Foot,
Phys. Rev. D \textbf{97}, 103006 (2018)
[arXiv:1801.09359];
R.~Foot,
Phys. Rev. D \textbf{97}, 043012 (2018)
[arXiv:1707.02528].




\bibitem{s2h}
R.~Foot,
Phys. Rev. D \textbf{98}, 123015 (2018)
[arXiv:1804.02847].

\bibitem{fv}
R.~Foot and R.~R.~Volkas,
Phys. Rev. D \textbf{70}, 123508 (2004)
[arXiv:astro-ph/0407522].


\bibitem{freview}
R.~Foot,
Int. J. Mod. Phys. A \textbf{29}, 1430013 (2014)
[arXiv:1401.3965].


\bibitem{paulo}
P.~Ciarcelluti and R.~Foot,
Phys. Lett. B \textbf{690}, 462-465 (2010)
[arXiv:1003.0880].


\bibitem{foots}
R.~Foot,
JCAP \textbf{04}, 014 (2012)
[arXiv:1110.2908];
R.~Foot and S.~Vagnozzi,
Phys. Lett. B \textbf{748}, 61-66 (2015)
[arXiv:1412.0762].


\bibitem{damares}
R.~Bernabei \textit{et al.} [DAMA],
Nucl. Instrum. Meth. A \textbf{592}, 297-315 (2008)
[arXiv:0804.2738].




\bibitem{range}
K.~Kannike, M.~Raidal, H.~Veermae, A.~Strumia and D.~Teresi,
[arXiv:2006.10735];
F.~Takahashi, M.~Yamada and W.~Yin,
[arXiv:2006.10035];
J.~Sun and X.~G.~He,
[arXiv:2006.16931];
C.~Cai, H.~H.~Zhang, G.~Cacciapaglia, M.~Rosenlyst and M.~T.~Frandsen,
[arXiv:2006.16267];
I.~M.~Bloch, A.~Caputo, R.~Essig, D.~Redigolo, M.~Sholapurkar and T.~Volansky,
[arXiv:2006.14521];
C.~Gao, J.~Liu, L.~T.~Wang, X.~P.~Wang, W.~Xue and Y.~M.~Zhong,
[arXiv:2006.14598];
P.~Athron {\it et al.},
[arXiv:2007.05517];
A.~N.~Khan,
[arXiv:2006.12887];
A.~Bally, S.~Jana and A.~Trautner,
[arXiv:2006.11919];
C.~Boehm, D.~G.~Cerdeno, M.~Fairbairn, P.~A.~N.~Machado and A.~C.~Vincent,
[arXiv:2006.11250]; 
O.~G.~Miranda, D.~K.~Papoulias, M.~Tortola and J.~W.~F.~Valle,
[arXiv:2007.01765];
U.~K.~Dey, T.~N.~Maity and T.~S.~Ray,
[arXiv:2006.12529];
B.~Fornal, P.~Sandick, J.~Shu, M.~Su and Y.~Zhao,
[arXiv:2006.11264];
N.~F.~Bell, J.~B.~Dent, B.~Dutta, S.~Ghosh, J.~Kumar and J.~L.~Newstead,
[arXiv:2006.12461];
G.~Choi, M.~Suzuki and T.~T.~Yanagida,
[arXiv:2006.12348]; 
H.~J.~He, Y.~C.~Wang and J.~Zheng,
[arXiv:2007.04963];
D.~Borah, S.~Mahapatra, D.~Nanda and N.~Sahu,
[arXiv:2007.10754];
S.~Karmakar and S.~Pandey,
[arXiv:2007.11892];
G.~Arcadi, A.~Bally, F.~Goertz, K.~Tame-Narvaez, V.~Tenorth and S.~Vogl,
[arXiv:2007.08500];
D.~McKeen, M.~Pospelov and N.~Raj,
[arXiv:2006.15140].


\bibitem{pole}
M.~Szydagis, C.~Levy, G.~M.~Blockinger, A.~Kamaha, N.~Parveen and G.~R.~C.~Rischbieter,
[arXiv:2007.00528].


\bibitem{md1}
L.~Zu, G.~W.~Yuan, L.~Feng and Y.~Z.~Fan,
[arXiv:2006.14577].



\bibitem{md2}
L.~Zu, R.~Foot, Y.~Z.~Fan and L.~Feng,
[arXiv:2007.15191].

\bibitem{xenonres}
E.~Aprile \textit{et al.} [XENON],
[arXiv:2003.03825].






\bibitem{sor}
P.~Sorensen,
Nucl. Instrum. Meth. A \textbf{635}, 41-43 (2011)
[arXiv:1102.2865].


\bibitem{s2100}
E.~Aprile \textit{et al.} [XENON],
Phys. Rev. D \textbf{94}, 092001 (2016)
[arXiv:1605.06262].

\bibitem{abc}
E.~Aprile \textit{et al.} [XENON],
Phys. Rev. D \textbf{100}, 052014 (2019)
[arXiv:1906.04717].

\bibitem{supx}
See Supplemental Material at http://link.aps.org/
\newline
\noindent
supplemental/10.1103/PhysRevLett.123.251801.


\bibitem{pan1}
E.~Aprile \textit{et al.} [XENON],
[arXiv:2007.08796].

\bibitem{pan2}
D.~S.~Akerib \textit{et al.} [LUX-ZEPLIN],
Phys. Rev. D \textbf{101}, 052002 (2020)
[arXiv:1802.06039].


\bibitem{pan3}
H.~Zhang \textit{et al.} [PandaX],
Sci. China Phys. Mech. Astron. \textbf{62}, 31011 (2019)
[arXiv:1806.02229].



\end{thebibliography}
\end{document}